\documentclass{article}

\usepackage{amssymb}
\usepackage{amsmath}

\begin{document}
\baselineskip18pt
\title{Entropy Methods in Random Motion}
\author{Piotr Garbaczewski\thanks{Presented at the XVIII Marian Smoluchowski
Symposium on Statistical  Physics, Zakopane, Poland, September 3-6, 2005}\\
Institute of Physics,  University  of Zielona  G\'{o}ra,
65-516 Zielona G\'{o}ra, Poland }%\\
%{ \it  Dedicated to Professor Peter Talkner on his 60th birthday}}
\maketitle
\begin{abstract}
We analyze  a contrasting dynamical behavior of Gibbs-Shannon  and conditional
Kullback-Leibler entropies, induced  by  time-evolution of
continuous probability distributions. The question  of  predominantly  \it  purpose-dependent \rm
  entropy definition for non-equilibrium model systems is addressed. The conditional
 Kullback-Leibler entropy is often  believed  to properly capture physical features of
 an asymptotic  approach towards   equilibrium.   We give arguments  in favor of  the  usefulness
 of   the standard  Gibbs-type entropy and indicate that its dynamics  gives an insight into
 physically relevant,  but  generally  ignored in the literature,  non-equilibrium phenomena.
 The role of physical units in the Gibbs-Shannon  entropy definition is discussed.
\end{abstract}
\vskip0.2cm
\hspace*{0.6cm} PACS numbers: 02.50.-r, 89.70.+c, 05.40.-a

\vskip0.2cm
\section{Introduction}

There are many notions of entropy. Except for the Clausius (thermodynamic) entropy, none of them  may be
considered unambiguously defined  or  to share  the  status of a physically  universal quantity  in the class
of dynamical  systems and phenomena, to the description of which a particular entropy notion has been possibly
designed.

Let us reproduce the standard (albeit non-exhaustive) list of entropies. For classical dynamical
 systems one is tempted to use any of:    Boltzmann, Gibbs, Shannon, Kullback-Leibler, Renyi,
 Tsallis, information/differential, topological, measure-theoretic and  Kolmogorov-Sinai
 entropies. In the quantum case  one encounters  von Neumann, Wehrl  and  Leipnik entropies, plus more
  or less natural/obvious generalizations of, classical by provenance, Kullback-Leibler, Renyi and
  Tsallis  entropies.
The concrete  entropy  choice is  with no doubt the context  (classical or quantum setting,
specific model system, specific notion of state, microstate and macrostate) and purpose-dependent.

We shall follow associations  born by non-equilibrium statistical physics  phenomena, where in the
 time-dependent problems such issues like  "trends" (convergence or  divergence)    towards
stationary  states plus Boltzmann-type  theorems (temporal behavior of H-functionals), validity, limitations,
possible violations, general rules of entropy evolution, meaning of the  entropy "production"/dissipation
and  its temporal behavior.

The term  \it entropy methods \rm  essentially refers to the mathematically rigorous discussion of the
asymptotic (large time) behavior   of   solutions of various partial
differential equations, in particular to these governing the  dynamics of  probability densities.
One attempts to quantify the  speed of con(div)ergence of  measures that allow to differentiate among
different solutions and their possibly different  temporal properties.

To set the stage to the main theme of our considerations,  let us invoke the
 simplest (naive) version of the Boltzmann   H-theorem, valid in case of the   rarified gas
  (mass $m$ particles),
without external forces, close to its thermal equilibrium,  under an assumption of its  space
homogeneity, \cite{huang,cercignani}. \\
  If the probability density function $f(v)$ is a solution of the  corresponding Boltzmann kinetic equation,
  then the  Boltzmann H-function (which coincides with  the negative of the   Gibbs-Shannon entropy)
   $H(t) = \int f(v) \ln f(v) dv $  does not increase:
\begin{equation}
{\frac{d}{dt}} H(t) \leq 0 \, . \label{boltz}
\end{equation}
 In particular, we know that there exists an
  invariant (asymptotic) density $ f_*(v) \simeq
   \exp[- m (v-v_0)^2/2k_BT]$ and  $H(t)$  is a constant  only if $f\doteq f_*(v)$.

Notice that in the one-dimensional  case, the  $L^1(R)$ density normalization coefficient
   reads $(m/2\pi k_BT)^{1/2}$ and thence, \it  formally, \rm    $H_* = \int f_* \ln f_* dv = - (1/2)
   \ln (2\pi e k_BT/m)$ where $e$ is the base of the natural logarithm. One must be aware of
      an apparent dimensional difficulty, \cite{bril}, since an argument of the logarithm is
      not  dimensionless.

Clearly, a  consistent integration outcome for $H(t)$  should involve a dimensionless argument
$k_BT/m[v]^2$ instead of $k_BT/m$,     provided  $[v]$ stands for \it  any \rm unit of velocity.
Examples are $[v]= 1\, m/s$    (here $m$ stands for the SI length unit, and not  for a mass parameter)
 or $10^{-5}\, m/s$.
To this end,  it suffices to redefine     $H_*$ as follows, \cite{bril,ohya}:
\begin{equation}
    H_* \rightarrow H_*^{[v]}=   \int f_* \ln ([v]\cdot  f_*) dv \, .
\end{equation}
Multiplying $f_*$ by $[v]$ we arrive at the  dimensionless argument of the logarithm in the above.

 We shall come back later to  a deeper  discussion of an impact of    dimensional
    units  on the  general  definition of  the Gibbs-Shannon  entropy
\begin{equation}
 {\cal{S}}(\rho ) = - \int  \rho (x) \ln \rho (x)\, dx
\end{equation}
  for $\rho \in L^1(R^n)$.

The entropy methods basically  refer to  the large time asymptotic  of the   heat and  Fokker-Planck
 equations, where in a mathematically oriented research all dimensional units, for the sake of clarity,
 are scaled away.
Following  \cite{toscani}, let us consider the heat equation in the re-scaled (no physical constants) form:
  $\partial _t u = \Delta u$ with  $x\in R^n$, $t\in R^+$ and $u(.,t=0)= u_0(.) \geq 0$,
$\int u_0(x) dx = 1$.

As $t \rightarrow \infty $, for any $u(x,t)$  we have $u(x,t) \simeq \rho (x,t) =
(4\pi t)^{-n/2} \exp [-x^2/4t]$, in conformity with the standard wisdom \cite{risken}
that  a regular  solution of the  heat equation behaves asymptotically as a
 fundamental solution,   once time goes to infinity.

There is a natural question to be addressed:
 what is  the $t\rightarrow \infty $  rate of convergence of the so-called Kullback "distance"
\begin{equation}
 \|u - \rho \|_{L^1}(t)
 \doteq \int |u(x,t) -\rho  (x,t)| dx
 \end{equation}
between two densities.  Since,    for  two  density functions $\rho $ and $\rho '$  there holds the
 Csisz\'{a}r-Kullback inequality, \cite{arnold}:
 \begin{equation}
  \int \rho  \ln (\rho /\rho ') dx \geq (1/2) \| \rho - \rho '\|^2_{L^1} \, ,
\end{equation}
it is the Kullback-Leibler entropy
\begin{equation}
{\cal{K}}(\rho ,\rho ') \doteq     \int \rho (x)\, \ln {\frac{\rho
 (x)}{\rho '(x)}}\, dx \, .
\end{equation}
which actually  stands for  an upper bound upon a
 "distance  measure" in the set of density functions.

If we consider $\rho _t$ to be a solution of the heat
 equation with the initial  data  $\rho _0$  and take  $ \rho
 _{\alpha}
(x)= (1/\sqrt{2\alpha \pi })\, \exp [- x^2/2\alpha ]$, then we may
always find $\alpha $ and $k$  such that $\rho _{\alpha + kt}$ has
the same second moment as $\rho _t$. This implies  an asymptotic $1/t$  decay of the
 initially  prescribed   Kullback-Leibler "distance",  \cite{toscani},
\begin{equation}
 {\cal{K}}(\rho
_t,\rho _{\alpha + kt}) \leq {\cal{K}}(\rho _0,\rho _{\alpha }) \label{largescale}
[\alpha /(\alpha + kt)] \,  .
\end{equation}

In view of the concavity of the function $f(w) = - w \ln
w$,  the Kullback-Leibler entropy  is  positive. This property if often contrasted with the fact
the  Gibbs-Shannon entropy ${\cal{S}}(\rho )$ may take negative values. Therefore, right at this point
(anticipating further discussion)  we  introduce the \it  conditional \rm
 Kullback-Leibler entropy  notion, which although non-positive by construction:
\begin{equation}
{\cal{H}}_c(\rho ,\rho ')   \doteq
- {\cal{K}}(\rho ,\rho ') \, ,
\end{equation}
 is  nonetheless  one of the major tools in the study of an asymptotic  convergence towards
an invariant   (equilibrium) density, \cite{tyran,tyran1}. This entropy  typically displays a prototype
behavior (monotonic growth in time), expected  to hold true if the  entropy definition is to be
compatible with  the casual understanding of the  second law of thermodynamics, \cite{tyran1}.

Now, let us consider the drifted Fokker-Planck (Smoluchowski) equation
 $\partial _t f = \Delta f - \nabla \cdot (b f)$, where $f(.,t) = f_0\geq 0$,
$\int f_0(x) dx =1$. We assume that the forward drift  $b=b(x,t)$ has a gradient form.
Let $f_*$ be the stationary solution of the F-P equation, then  an obvious question is:
 what is  the $t\rightarrow \infty $  rate of convergence of
$\|f - f_* \|_{L^1}(t) \doteq \int |f(x,t) - f_*(x)| dx$  towards the value  $0$  ?

 The outcome, albeit not completely general, is
that $\rho _t$ decays   \it in relative entropy \rm   to a
Gaussian(Maxwellian), the  speed of such  decay is exponential,
\cite{arnold}.   This is typically encoded in the formula, \cite{arnold,tyran,tyran1}
  of the form
\begin{equation}
 {\cal{H}}_c(t) \simeq \exp (-\alpha t) {\cal{H}}_c(0) \, , \label{decay}
\end{equation}
where   ${\cal{H}}_c(t) \doteq
{\cal{H}}_c(f_t,f_*)$,  with   $\alpha >0$ and  $f_t\doteq f(x,t)$, $t\geq 0$.
See also an explicit   discussion of the Ornstein-Uhlenbeck process in \cite{gar}.

In the course of  the time evolution, the conditional entropy monotonically  approaches
its maximum at zero, \cite{tyran1}. This property is seldom shared by the Gibbs-Shannon entropy of the
involved  time-dependent  probability density. The Gibbs entropy may grow, diminish,
oscillate and show more complicated patterns of behavior, \cite{tyran1,gar,gar1}.
 A physical relevance of  such  "strange" temporal  properties,  compare e.g.
  Eq.~(\ref{boltz}),  is worth  addressing   and it is our main goal in the present paper.

\section{Gibbs-Shannon and Kullback-Leibler entropies}

A casual understanding of  the entropy notion in  physics  is that  entropy  (tacitly one presumes to deal with
its  thermodynamic Clausius  version) is  a measure of  the degree of randomness and the  tendency (trend)
 of physical systems to become less and less organized.
We attribute  a  very concrete  meaning to the term  \it organization \rm  -  namely, we are interested in
quantifying   how good is the  probability  \it  localization \rm  on the state space (whatever: configuration
space, velocity or phase-space) of the system.

As a hint  let us consider   a probability measure $\mu = (\mu _1,\mu _2,...,\mu _N)$ on a system
of $N$ points, e. g.  $\sum_{j=1}^N \mu _j = 1$.  The standard Shannon entropy reads
$S(\mu ) = - \sum_{j=1}^N  \mu _j \log \mu _j \Longrightarrow
0\leq S(\mu ) \leq \log N$  and its  maximum of corresponds to a  uniform  probability distribution
$\mu _j =1/N$   for all $j$.

  If $X$ is a  discrete random variable  taking values $x_i$ with probabilities $p_i$, $i=1,2,...,N$,
  the quantity
${\cal{S}}(X) = - \sum p_i \log p_i $ is  called the  Shannon entropy of a discrete random variable  or
 the entropy of the probability distribution $(p_1,...,p_N)$. If $X$ takes infinitely many values
  $x_1,x_2,...$ with probabilities $p_1,p_2,...$,  then the entropy
${\cal{S}}(X)$ is not necessarily finite.

As a side comment we recall that $\log $ has base $2$ in which case  the unit of entropy is called  a  bit
 (binary digit), while for $\ln $ with  base $e$, the unit of entropy is called  a  nat (natural); we
   observe that    $\log  b\, \cdot \ln 2 = \ln b$.

For a   continuous random variable $X$  with values in $x\in R^n$ and  the probability density $\rho (x)$
one usually defines the  Shannon entropy of a continuous  random variable (called   the
differential entropy) of $X$) as: $ {\cal{S}}(X) = - \int_{\Gamma} \rho (x) \log \rho (x) dx$,
where $\Gamma \in R^n$ is the support set of $X$. One may also denote ${\cal{S}}(X) \doteq {\cal{S}}(\rho )$.

There is  number of standard   views about the discrete and continuous entropies.
In the  discrete case, the entropy quantifies randomness in  an \it absolute \rm  way.
In the continuous case there is no smooth limiting passage from the  discrete to continuous
entropy. Then, the entropy cannot  work  "as it is"   as a measure of global randomness and
one usually invokes  a  casual list of  drawbacks:  ${\cal{S}}(\rho )$ may
 be negative, may be  unbounded  both from below and above, is scaling (hence coordinate transformation)
   dependent.

Anyway, a  difference of two  Shannon  entropies,  necessarily evaluated with respect to the
 same coordinate system,
 $ {\cal{S}}(\rho ) -  {\cal{S}}(\rho ')$ is known
to  quantify an  absolute change in the  information/randomness   content  when  passing
from  $\rho $ to $\rho '$ and is obviously scaling independent. The same observation
extends to the time derivative of the Shannon entropy in case of time-dependent probability densities.

Alternatively, although with reservations,   one  may  pass to  the
 familiar notion of the  Kullback-Leibler    entropy ${\cal{K}} =\int_{\Gamma } \rho \, (\ln
{\rho } - \ln {\rho '})\, dx$, non-negative and scaling-independent from the outset.
However, one  should keep in mind that it is the conditional Kullback-Leibler (K-L)
 entropy ${\cal{H}}_c = - {\cal{K}}$ which is predominantly used in the literature as a justification,
  in terms of model systems,  of the "entropy growth paradigm". Like ${\cal{S}}(\rho )$,
  the conditional K-L  entropy takes negative  values  and  its upper bound actually  equals zero.

  Let us point out that a  consistent exploitation of the conditional  K-L entropy  is restricted either to
  the large time-scale phenomena, see e.g.  Eq.~(\ref{largescale}), or to the dynamical systems which have
  an invariant density, see Eq.~(\ref{decay}).
  In the short time-scale regimes and  for  systems without invariant densities,   the conditional
  Kullback-Leibler entropy is not an adequate tool.

  Let us  consider
\begin{equation}
\rho _{\alpha, \beta }
 = \beta \, \rho  [\beta (x-\alpha )] \, . \label{scale}
 \end{equation}
 where $\alpha \geq 0, \beta >0$ are  real  parameters.
The respective Shannon entropy reads:
\begin{equation}
{\cal{S}}(\rho _{\alpha ,\beta })  = {\cal{S}}(\rho )  -
 \ln \beta  \, .
\end{equation}
For   general  probability distributions $\rho (x)$ with a  fixed variance $\sigma $
we  have $S(\rho )\leq  {\frac{1}2}  \ln (2\pi e \sigma ^2)$  and
$S(\rho )$  becomes maximized if and only if $\rho $ is a Gaussian.
Therefore we can write
\begin{equation}
(2\pi e)^{-1/2}\,
  \exp [{\cal{S}}(\rho _{\alpha ,\beta })]  \leq     \sigma /\beta
\end{equation}
and give a meaning to   the $\beta $-scaling transformation of $\rho (x-\alpha ) $:
 the density  is broadened  if  $\beta <1$  and   shrinks   if  $\beta >1$.

Given  a one parameter family
 of  Gaussian densities   $\rho _{\alpha  } = \rho (x-\alpha  )$, with the mean  $\alpha  \in R$ and
 the  standard deviation fixed at $\sigma $.  These densities  share   the
 very same value of Shannon entropy, independent of $\alpha $:
 $$ {\cal{S}}_{\sigma }
 =   {\frac{1}{2}} \ln \,(2\pi e \sigma ^2)$$.

If we admit the standard deviation  $\sigma $ to be another free parameter, a two-parameter family
  $\rho _{\alpha  } \rightarrow \rho _{\alpha , \sigma }(x)$ appears. Then:
$$
{\cal{S}}_{\sigma '} - {\cal{S}}_{\sigma } =  \ln \, \left(
\frac{\sigma '}{{\sigma }}\right) \, .  \label{comparison}
$$

By denoting    $\sigma \doteq \sigma (t) = \sqrt{2Dt}$ and $\sigma ' \doteq \sigma (t')$
we  make  the non-stationary (heat kernel)  density  amenable to the "absolute comparison"  formula
  at different time instants  $t'>t >0$: $(\sigma '/ \sigma ) = \sqrt{t'/t}$.

Indeed a fundamental solution of the heat equation
$\partial _t \rho = D \Delta \rho $ reads
\begin{equation}
\rho (x,t) = {\frac{1}{(4\pi Dt)^{1/2}}} \exp \left( - {\frac{x^2}{4Dt}}\right)
\end{equation}
 whose  differential entropy equals ${\cal{S}}(t) =  (1/2) \ln (4\pi e Dt) $, or in the
  dimensionless form:  ${\cal{S}}^{[x]}(t) =
 (1/2) \ln (4\pi e Dt/[x]^2) $, where $[x]$ is any dimensional  unit with the  SI  dimension
  of length.

Let $\rho _{\upsilon }$ denote  a convolution of a probability density  $\rho $
 with  a Gaussian
probability density having variance $\upsilon $.   The transition density (heat kernel) of
the  Wiener process generates such a convolution for any $\rho _0(x)$,
 with $\upsilon = \sigma ^2
\doteq 2Dt$.   Then, (de Bruijn) we have the entropy accumulation  formula:
$$
{\frac{d{\cal{S}}}{dt}} =  D \cdot{\cal{F}}  =    D\cdot \int
{\frac{(\nabla  \rho )^2}{\rho }} dx > 0
$$
 The monotonic  growth of ${\cal{S}}(t)$  is paralleled by   linear in time  growth  of
 the standard  deviation   $\sigma (t)$, hence quantifies  the  uncertainty (disorder) increase related
to  the "flattening" down of $\rho $.

Let us consider the  Kullback  entropy  ${\cal{K}}(\theta ,\theta')$ for
a  family of probability densities $\rho _{\theta}$ labelled by a parameter (one or more) $\theta $,
so that the   "distance" between  any two densities in this family
can be directly  evaluated.  We take   $\rho _{\theta '}$ as
reference probability density. Then:
\begin{equation}
{\cal{K}}(\theta ,\theta ') \doteq  {\cal{K}}(\rho _{\theta }|\rho _{\theta '})
=    \int \rho _{\theta }(x)\, \ln
{\frac{\rho _{\theta }(x)}{\rho _{\theta '}(x)}}\, dx \, .
\end{equation}

It is particularly instructive to evaluate various  K-L -  "distances" among members of a two-parameter
 family of  $L^1(R)$-normalized  Gaussian functions,  labelled  by  independent
parameters $\theta _1 = \alpha $ and $\theta _2 = \sigma $
(alternatively $\theta _2 = \sigma ^2$)  such that $\theta \doteq (\theta _1,\theta _2)$.
In the self-explanatory notation, for  two different   $\theta $ and $\theta '$
Gaussian densities there holds:
\begin{equation}
{\cal{K}}(\theta ,\theta ')=  \ln {\frac{\sigma '}{\sigma }} +
{\frac{1}{2}}({\frac{\sigma ^2}{{\sigma '}^2}} - 1) +
{\frac{1}{2{\sigma '}^2}} (\alpha - \alpha ')^2  \, .
\end{equation}
We may assume that $\theta '$  very little deviates from  $\theta $: $\theta ' = \theta + \Delta \theta $.
Then, we have
\begin{equation}
{\cal{K}}(\theta ,\theta + \Delta  \theta ) \simeq {\frac{1}{2}}
 \sum_{i,j} {\cal{F}}_{ij}\, \cdot  \Delta \theta _i  \Delta \theta _j
\end{equation}
where $i,j, = 1,2$ and the   Fisher information matrix
${\cal{F}}_{ij}$ has the form
\begin{equation}
{\cal{F}}_{ij} = \int \rho _{\theta } {\frac{\partial \ln \rho
_{\theta } }{\partial \theta _i}}  \cdot {\frac{\partial \ln \rho
_{\theta } }{\partial \theta _j}}\, dx \, .
\end{equation}
In case of Gaussian densities, labelled  by  independent  $\theta _1= \alpha , \theta _2 = \sigma $ (or
$\theta _2= \sigma ^2 $) the Fisher matrix is diagonal.

Let us set $\alpha ' = \alpha $ and consider $\sigma ^2 =
2Dt$, $\Delta (\sigma ^2)= 2D\Delta t$. Then ${\cal{S}}(\sigma '^2)
- {\cal{S}}(\sigma ^2) \simeq \Delta t/2t$, while ${\cal{K}}(\theta
,\theta ') \simeq (\Delta t)^2/ 4t^2$. Although, for finite
increments $\Delta t$ we have
$$
 {\cal{S}}(\sigma '^2) -
{\cal{S}}(\sigma ^2)\simeq \sqrt{ {\cal{K}}(\theta ,\theta ')}\simeq
{\frac{\Delta t}{2t}} \, ,
$$
 the time derivative
 notion $\dot{\cal{S}}$  surely  can be   defined   for the differential entropy,
 but is  definitely meaningless  in terms of  the corresponding short time-scale
  Kullback "distance", c.f. \cite{gar,gar1}.

We stress that  no such obstacle arises  in the standard cautious  use of the
 conditional   Kullback entropy ${\cal{H}}_c$, when an invariant density is in hands.  Indeed, normally one
 of the involved densities  is the  stationary  (reference) one
 $\rho _{\theta '}(x) \doteq \rho _*(x)$,  while    another is allowed to  evolve in time
 $\rho _{\theta }(x) \doteq \rho (x,t)$, $t\in R^+$, thence
 ${\cal{H}}_c(t) \doteq - {\cal{K}}(\rho _t|\rho _*)$ and $d{\cal{H}}_c(t)/dt$ does make sense.

We recall that for the free Brownian motion there is no invariant density.
As we have indicated before, Eq.(\ref{largescale}),    ${\cal{H}}_c(\rho _t,\rho _{t'})$, $t<t'$
still remains a useful tool, albeit in the asymptotic regime and for not too small values of $t'-t$.

\section{Physical  units in the entropy definition}

Let us come back to an issue of physical units in the definition of a differential entropy.
 In fact, if $x$ and $p$ stand for
one-dimensional phase space labels and $f(x,p)$ is a normalized
phase-space density, $\int  f(x,p) dx dp=1 $, then the related \it
dimensionless \rm differential entropy reads  as follows,
\cite{ohya}:
\begin{equation}
{\cal{S}}_h = - \int (h f) \ln (h  f) {\frac{dx dp}h}  = - \int f
\ln (h f) dx dp
\end{equation}
where $h=2\pi \hbar $ is the  tentatively  accepted (there is no other  mention of quantum theory)
 Planck constant. Let  $\rho  (x)$  and $\tilde{\rho }_h(p)$  be two independent , respectively
 spatial and momentum space  densities. We form  the joint density
\begin{equation}
f(x,p) \doteq \rho (x) \tilde{\rho }_h(p)
\end{equation}
and evaluate the differential entropy   ${\cal{S}}_h$ for this
density.  Remembering that  $\int \rho  (x) dx =1=  \int
\tilde{\rho }_h(p) dp$, we have formally:
\begin{equation}
{\cal{S}}_h = - \int  \rho \ln \rho dx - \int \tilde{\rho }_h \ln
\tilde{\rho }_h\, dp  - \ln h = S^x + S^p  - \ln h  \, . \label{deco}
\end{equation}
 The formal use of the logarithm properties before executing integrations in  $\int
\tilde{\rho }_h \ln (h \tilde{\rho }_h)\, dp$, has left us with
an  issue  of "literally taking the logarithm of a dimensional argument" i. e.  that of
$\ln h$.

We recall that ${\cal{S}}_h $ is a dimensionless quantity, while
if $x$ has dimensions of length,     then the probability density
has dimensions of inverse length and  analogously in connection
with momentum dimensions.

Let us denote $x\doteq   r \delta x$ and $p\doteq \tilde{r} \delta
p$ where labels $r$ and $\tilde{r}$ are dimensionless, while
$\delta x$ and $\delta p$ stand  for respective position and
momentum  dimensional (hitherto - resolution) units. Then:
\begin{equation}
 -\int \rho  \ln
\rho dx  - \ln (\delta x)  \doteq  -\int \rho \ln (\delta x \rho
)dx  \label{34}
\end{equation}
 is a dimensionless quantity. Analogously
\begin{equation}
-\int \tilde{\rho }_h \ln \tilde{\rho }_h\, dp - \ln \delta p
\doteq -\int \tilde{\rho }_h \ln (\delta p \tilde{\rho }_h)\, dp
\end{equation}
is  dimensionless. First left-hand-side terms in two above
equations we recognize as $S^x$ and $S^p$ respectively.

Hence, formally we have arrived at  a manifestly dimensionless decomposition
\begin{equation}
{\cal{S}}_h = -\int \rho \ln (\delta x \rho )dx  -\int \tilde{\rho
 }_h \ln (\delta p \tilde{\rho }_h)\, dp  + \ln {\frac{\delta x
\delta p}h} \doteq S^x_{\delta x} + S^p_{\delta p} +\ln
{\frac{\delta x \delta p}h} \label{final}
\end{equation}
instead of the previous one, Eq.~(\ref{deco}). The last identity
Eq.~(\ref{final})  gives an unambiguous meaning to the preceding
formal  manipulations with  dimensional quantities. Instead of the Planck
constant $h$ we can use any other unit with  SI dimensions of action,
 say $\delta h$.

 As a byproduct of our discussion,
we have resolved the case of the spatially interpreted real axis,
when $x$ has dimensions of length, c.f. also \cite{ohya}:
$S^x_{\delta x} = -\int \rho \ln (\delta x \rho )dx$ is the pertinent
 dimensionless differential entropy definition for spatial probability densities.

{\bf Example 1:} Let us discuss an explicit example involving
the Gauss density
\begin{equation}
 \rho (x)= (1/\sigma \sqrt{2\pi })\, \exp [-
(x-x_0)^2/2\sigma ^2]   \label{gaussian}
\end{equation}
 where $\sigma $ is the standard deviation
(its square stands for the variance). There holds ${\cal{S}}(\rho
) =  {\frac{1}{2}} \ln \,(2\pi e \sigma ^2)$ which is  a
dimensionless outcome.  If we pass to $x$ with dimensions of
length, then  inevitably $\sigma $ must have dimensions of length.
It is instructive to check that in this dimensional case we have a
correct dimensionless result:
\begin{equation}
S^x_{\delta x} = {\frac{1}{2}} \ln \,[2\pi e \left({\frac{\sigma
}{\delta x}}\right)^2]
\end{equation}
to be compared with Eq.~(\ref{34}).  Clearly,
$S^x_{\delta x}$ vanishes if $\sigma /\delta x = (2\pi e )^{-1/2}
$, hence at the dimensional value  of the standard deviation
$\sigma = (2\pi e )^{-1/2} \delta x$, compare e.g. \cite{ohya}.

{\bf Example 2:} Let us invoke the
 simplest (naive) text-book  version of the Boltzmann   H-theorem, valid in case of the   rarified gas
  (of mass $m$ particles), without external forces, close to its thermal equilibrium,  under an
assumption of its  space homogeneity, \cite{huang,cercignani}.
  If the probability density function $f(v)$ is a solution of the corresponding  Boltzmann
  kinetic equation, then the Boltzmann $H$-functional  (which is
  simply the negative of the differential entropy)    $H(t) = \int f(v) \ln f(v) dv $  does not increase:
  ${\frac{d}{dt}} H(t) \leq 0 $.
 In the present case  we know that there exists an
  invariant (asymptotic) density, which in one-dimensional case has the form  $ f_*(v)
  = (m/2\pi k_BT)^{1/2}  \exp[- m (v-v_0)^2/2k_BT]$.    $H(t)$  is  known to be
   time-independent  only if $f\doteq f_*(v)$.
   We can straightforwardly   evaluate $H_* = \int f_* \ln f_* dv = - (1/2)
   \ln (2\pi e k_BT/m)$  and become  faced with  a  an apparent dimensional difficulty, \cite{bril}:
   an argument of the logarithm is not  dimensionless.
For sure, a   consistent     integration outcome for $H(t)$ should
involve   $k_BT/m[v]^2$ instead of $k_BT/m$,     provided  $[v]$ stands for \it  any \rm unit of
velocity. Examples are $[v]= 1\, m/s$     (here $m$ stands for the
SI  length unit, and not  for a mass parameter) or $10^{-5}\, m/s$.
To this end it suffices to redefine  $H_*$ as follows, \cite{bril,ohya}:
\begin{equation}
    H_* \rightarrow H_*^{[v]}=   \int f_* \ln ([v]\cdot  f_*) dv \, .
\end{equation}
Multiplying $f_*$ by $[v]$ we arrive at the dimensionless argument
of the logarithm in the above and cure the dimensional obstacle.

We   recall  that under the scaling transformation  Eq.~(\ref{scale})  the respective Shannon entropy
takes the form  ${\cal{S}}(\rho _{\alpha ,\beta })  = {\cal{S}}(\rho )  -
 \ln \beta  $.   In  case of  Gaussian $\rho $,  we get
${\cal{S}}(\rho _{\alpha ,\beta })=  \ln [(\sigma /\beta )
\sqrt{2\pi e}]$.
Clearly, ${\cal{S}}(\rho _{\alpha ,\beta
 })$ takes the value $0$ at $\sigma = (2\pi e )^{-1/2} \beta $ in analogy with our previous
 dimensional considerations. If an  argument of $\rho $ is assumed
 to have dimensions, then the scaling transformation with the
 dimensional $\beta $ may be interpreted as a method to restore
 the dimensionless differential entropy value.

\section{Temporal behavior of entropies}

\subsection{Deterministic system}

Let us  consider a classical dynamical system in $R^n$  whose evolution
is governed by equations of motion:
\begin{equation}
\dot{x} = f(x) \label{set}
\end{equation}
where $\dot{x}$ stands for the time derivative and $f$  is an  $R^n$-valued
function of $x\in R^n$,  $x= \{ x_1,x_2,...,x_n\}$.
A statistical ensemble of solutions  of such dynamical equations   can be
 described by a time-dependent probability density  $\rho (x,t)$ whose dynamics
 is given by the generalized  Liouville (in fact, continuity) equation
 \begin{equation}
\partial _t \rho = - \nabla \cdot (f\, \rho )
\end{equation}
where $\nabla \doteq \{ \partial /\partial x_1,...,\partial /\partial x_n\} $.

With a continuous probability density $\rho \doteq \rho(x,t) $,  where $x\in
R^n$ and  we allow for  an explicit time-dependence, we
associate a respective differential entropy functional  $ {\cal{S}}(\rho )$, where
in general  ${\cal{S}}(\rho ) \doteq  {\cal{S}}(t)$ depends on time.

 Let us  take for granted that
an interchange of time derivative  with  an indefinite  integral is allowed (suitable
precautions are necessary with respect to  the convergence of integrals). Then, we readily get
an identity:
\begin{equation}
\dot{\cal{S}} =  \int \rho \, (div \, f) dx \doteq \langle \nabla \cdot f \rangle \label{Liouville}\, .
\end{equation}
Accordingly, the information entropy ${\cal{S}}(t)$ grows with time only if the dynamical
system has  positive  mean   flow divergence.

However, in general  $\dot{\cal{S}}$ is  not positive definite.
For example, dissipative dynamical  systems  are characterized by the   negative
(mean) flow divergence.
Fairly often, the divergence of the flow  is constant. Then,  an "amount of
information" carried by a  corresponding  statistical ensemble (e.g. its density)  increases,  which is
 paralleled by  the information  entropy decay (decrease).

An example of a system with a point  attractor  (sink) at origin is a one-dimensional
 non-Hamiltonian  system $\dot{x}= - x$.  In this case  $div f = -1$ and $\dot{\cal{S}} = -1$.
 Further discussion  of dynamical systems  with strange (multifractal) attractors, for which
 the Shannon information  (differential) entropy  decreases indefinitely (the pertinent steady states are
 no longer represented by probability density functions) can be found in \cite{nicolis}.
 We note that for Hamiltonian systems,
  the phase-space flow has vanishing  divergence, hence $\dot{\cal{S}} =0$ which implies that "information
   is conserved" in Hamiltonian dynamics.

Let  there be given  an invertible  dynamical system on $R^2$,
 with  $f(x)\doteq  F x$, where $F$ is a
two-by two real  matrix and $x\in R^2$, \cite{tyran1}. A solution has the form  $x(t) = \exp(tF) x(0)$, where
the matrix operator $\exp(tF)$ is defined through the standard Taylor expansion formula. The solution
of the Liouville equation with  an initial probability density $f_0(x)$ is given by
\begin{equation}
f(x,t) = \exp [-(tr F) t]\cdot f_0(\exp(-tF)x)\, .
\end{equation}
and hence:
\begin{equation}
{\cal{S}}(f_t) = {\cal{S}}(f_0) +  (tr F) t \Rightarrow \dot{\cal{S}}(f_t) = tr F  \label{det}
\end{equation}
Obviously $Tr F= \lambda _1+\lambda _2$, where $\lambda _i, i=1,2$ are   the eigenvalues of $F$.
We realize that ${\cal{S}}(f_t)$ grows indefinitely if $tr F>0$ and diminishes  indefinitely towards $-\infty $
if $tr F <0$.  There is no stationary density  and  the conditional entropy is not defined.

\subsection{Random system}

In case of a general  dissipative dynamical system, a controlled  admixture
of noise  can stabilize dynamics and yield asymptotic invariant densities.
For example, an additive modification of
the right-hand-side of   Eq.~(\ref{set}) by white noise term $A(t)$ where
$\langle A_i(s)\rangle =0$ and $\langle A_i(s)A_j(s')\rangle = 2q \delta (s-s') \delta _{ij}$,
 $i=1,2,...n$, implies the Fokker-Planck-Kramers equation:
 \begin{equation}
\partial _t \rho = - \nabla \cdot (f\, \rho )  + q \Delta \rho
\end{equation}
where $\Delta \doteq \nabla ^2 = \sum_{i} \partial ^2  /\partial x_i^2$.
 Accordingly, the differential entropy dynamics  would take another form than this  defined by
  Eq.~(\ref{Liouville}):
\begin{equation}
\dot{\cal{S}} =  \int \rho \, (div \, f) dx   + q \int {\frac{1}{\rho }}  \label{daems}
 (\nabla \rho )^2\, dx   .
\end{equation}
Now, the     dissipative  term  $\langle \nabla \cdot f \rangle <0 $ can be counterbalanced
by a strictly positive    stabilizing contribution  $q \sum_{i}  \int {\frac{1}{\rho }}
 (\partial \rho /\partial x_i)^2\, dx$. This allows
  to expect that, under suitable circumstances  dissipative systems with noise may
  yield  $ \dot{\cal{S}} = 0$.  If   $\langle \nabla \cdot f \rangle \geq 0 $,
   then  the  differential (information)  entropy would    grow monotonically.

We shall discuss an example of a  non-invertible system,  provided by   the standard
one-dimensional Ornstein-Uhlenbeck process, \cite{gar,tyran}.
We choose  the  forward drift of the Fokker-Plack equation    $\partial _t\rho =
D\triangle \rho  +  {\nabla } [(\gamma  x)  \rho ]$   with  $\gamma >0$ and $D>0$
being  the diffusion coefficient.

If  an initial density is chosen in the Gaussian form, with the mean value $\alpha _0$ and variance $\sigma ^2_0$.
the  Fokker-Planck evolution  preserves the Gaussian form  of $\rho (x,t)$ while modifying
 the mean value
$\alpha (t) = \alpha _0 \exp(-\gamma t)$ and variance:
\begin{equation}
 \sigma ^2(t) = \sigma ^2_0 \exp (- 2 \gamma t) + {\frac{D}{\gamma }}[1-\exp (-2\gamma t)] \, .
\end{equation}

Accordingly, since   a unique invariant density has the form $\rho _* =   \sqrt{\gamma /2\pi D}
 \exp (-\gamma x^2/2D)$ we obtain:
\begin{equation}
{\cal{H}}_c(t) = \exp (-2 \gamma t) {\cal{H}}_c(\rho _0,\rho _*)= -{\frac{\gamma \alpha _0^2}{2D}}\,
\exp (-2\gamma t)
\end{equation}
i.e. a monotonic  growth of  the negative-valued  conditional Kullback-Leibler entropy towards
its maximum at zero:
\begin{equation}
\dot{\cal{H}} _c(t)=   - 2 \gamma  \exp (-2 \gamma t) {\cal{H}}_c(\rho _0,\rho _*)   =
\gamma ^2 {\frac{\alpha _0^2}{D}}\,
\exp (-2\gamma t) > 0  \, .  \label{grow}
\end{equation}
The  differential entropy:
\begin{equation}
{\cal{S}}(t) =(1/2) \ln [2\pi e \sigma ^2(t)]
\end{equation}
shows another temporal behavior
\begin{equation}
\dot{\cal{S}}= {\frac{2\gamma  (D - \gamma \sigma _0^2) \exp(-2 \gamma t)}
{D -(D - \gamma \sigma _0^2)\exp(-2\gamma t)}}\, .
\end{equation}
We observe that if $\sigma ^2_0 > D/\gamma $, then $\dot{\cal{S}} <0$, while $\sigma ^2_0 < D/\gamma $
implies $\dot{\cal{S}} > 0$.

In both cases the behavior of the differential   entropy is monotonic,
although its growth or decay do critically rely on  the choice of $\sigma ^2_0$.
Irrespective of $\sigma ^2_0$ the asymptotic value of   ${\cal{S}}(t)$ as
  $t\rightarrow \infty $  reads $(1/2) \ln [2\pi e (D/\gamma )]$.  It is useful to note,  that
  in the special case of $\sigma ^2_0 = D/\gamma $ the differential entropy is a constant of motion, while
  the conditional K-L entropy nonetheless  does grow,  asymptotically  approaching the value zero
     according to Eq.~(\ref{grow}).

Summarizing, we can  say   that  the conditional Kullback-Leibler entropy  of the
Ornstein-Uhlenbeck  process grows monotonically in time, while the temporal behavior of the
Gibbs-Shannon (differential) entropy  depends on statistical properties (half-width  $\sigma _0$)
of the initial ensemble density. This pattern of temporal behavior appears to   be generic to a large class of
 dynamical systems, \cite{tyran1}.

To find out whether there is anything deeper  in the above apparent differences  in the temporal behavior of
the Gibbs-Shannon and Kullback-Leibler entropies associated with the same time-dependent probability density,
except for the a priori presumed existence of the reference invariant density, let us consider  the
one-dimensional Fokker-Planck equation for any Smoluchowski process.  We assume
\begin{equation}
\partial _t\rho =
D\triangle \rho -  {\nabla } (b  \rho ) \label{Fokker}
\end{equation}
with a forward drift $b= b(x,t)$ of the gradient form $b= - \nabla \Phi $ and  attribute to
a  diffusion   coefficient $D$ dimensions of $\hbar /2m$ or $k_BT/m\beta $.

Furthermore, we introduce  the velocity fields: $u(x,t) = D \nabla \ln \rho (x,t) $
and $v(x,t) = b(x,t) - u(x,t)$. The current velocity $v(x,t)$, in view of
$\partial _t \rho = - \nabla (v \rho ) $ which is an equivalent form of Eq.~(\ref{Fokker}),
  contributes to  the diffusion current $j=v\rho $.

For the  differential  entropy
${\cal{S}}(t) = - \int \rho (x,t)\, \ln \rho  (x,t)\,  dx $, while   imposing
 boundary restrictions that $\rho, v\rho, b\rho $ vanish at spatial infinities or  finite
interval borders,   we readily get the entropy  balance equation of the form Eq.~(\ref{daems}),
with the minor modification i. e.  the replacement
of $q$ by $D$.  We are however interested in its equivalent form (easily derivable under previously listed
boundary restrictions), \cite{gar,gar1}:
\begin{equation}
D \dot{\cal{S}}  =  \left< {v}^2\right>
    -  \left\langle {b}\cdot {v}
 \right\rangle \, .
\end{equation}

Remembering that we deal with the Smoluchowski process, we set (adjusting dimensional constants):
$ b = (D/k_BT)\,  F$.
Exploiting  $j \doteq v\rho $ and  demanding  $ F = - \nabla V $ we infer:
\begin{equation}
\dot{\cal{S}}= (1/D) \left< v^2 \right>
  - \dot{\cal{Q}}
\end{equation}
where the first (positive)   term on the right-hand-side stands for the differential  entropy  accumulation rate
 (entropy gain by the system).

  The     second term contains the $\dot{\cal{Q}}$  entry:
\begin{equation}
\dot{\cal{Q}} \doteq   (1/k_BT) \int
{F} \cdot {j}\,  dx = (1/D) \left\langle {b}\cdot {v} \label{heat}
 \right\rangle \,
 \end{equation}
which , if positive ($\dot{\cal{Q}}>0$ is not a must, \cite{gar}), allows to  interpret $- \dot{\cal{Q}}$
 as the  entropy dissipation rate,  i.e. an entropy transfer to the environment  in the form of the surplus  heat.
Note that \ $k_BT \dot{{\cal{Q}}}= \int {F}\cdot {j}\,  dx$  has a conspicuous from of the
fairly standard   power release expression i.e. the time rate at which the mechanical work per unit
of mass  is returned back to the thermal reservoir (or absorbed if   $\dot{\cal{Q}}<0$) in the form of heat.

Under current premises, there exists a stationary solution of the Fokker-Planck equation
\begin{equation}
\rho _*(x)= {\frac{1}{Z}} \exp \left(-{\frac{V(x)}{k_BT}}\right)  \label{inv}
\end{equation}
 where $Z=\int \exp(-V(x)/k_BT)\, dx $.

Let us take  $\rho _*(x)$  as a reference density  with respect to which the divergence of $\rho (x,t)$
is quantified in terms of the conditional K-L entropy. Then:
\begin{equation}
{\cal{H}}_c(t) =  -  \int     \rho \,
\ln \left({\frac{\rho }{\rho _*}}\right )\, dx  =  {\cal{S}}(t) - \ln Z - {\frac{\langle  V\rangle }{k_BT}}
\end{equation}
and   straightforwardly,  because of
\begin{equation}
{\frac{d}{dt}} \left< V \right>
 = - k_BT \dot{\cal{Q}}   \label{heat1}
\end{equation}
we arrive at
\begin{equation}
 \dot{\cal{H}}_c = \dot{\cal{S}} + \dot{\cal{Q}}  \geq 0  \, .
\end{equation}

At this point, we can come back to a continued discussion of the Ornstein-Uhlenbeck process.
Namely, we have here a direct control of the behavior of the "power release" expression $\dot{\cal{Q}} =
\dot{\cal{H}}_c - \dot{\cal{S}}$.
 Since
\begin{equation}
\dot{\cal{H}}_c= (\gamma ^2\alpha _0^2/D) \exp(-2\gamma t) >0\, ,
\end{equation}
in case of $\dot{\cal{S}} <0$ we encounter
 a continual power supply  $\dot{\cal{Q}}> 0$  by the thermal environment (alternatively, power absorption by
 the system).

 In case of $\dot{\cal{S}} >0$ the situation is more complicated. For example, if $\alpha _0 =0$, we
  can easily check that  $\dot{\cal{Q}}  < 0$, i.e. we have the power drainage from the environment  for all $t\in R^+$.
  More generally, the sign of  $\dot{\cal{Q}}$ is
  negative for   $\alpha _0^2< 2(D-\gamma \sigma ^2_0)/\gamma$. If the latter inequality  is reversed,
  the sign of $\dot{\cal{Q}}$ is not uniquely specified  and suffers  a change at a suitable time  instant $t_{change}
  (\alpha _0^2,\sigma _0^2)$.

Interestingly enough, in the special case of  $\sigma ^2_0= D/ \gamma $ i. e. $\dot{\cal{S}}=0$, we encounter
\begin{equation}
\dot{\cal{H}}_c =  \dot{\cal{Q}}  \geq 0
\end{equation}
 i.e. a direct connection between the entropy increase    and heat  removal (to the thermostat)
 \it  time rates, \rm  which counterbalance each other.

\subsection{Phase-space dynamics}

One may argue that the reported above, rather unexpected,  insight into the nontrivial power transfer processes
is an artifact of the one-dimensional spatial (Smoluchowski) projection of the phase-space  motion.
Let us therefore indicate arguments to the contrary.

For Hamiltonian systems the phase-space flow is divergence-less. Indeed, let us consider  a two-dimensional
conservative system $\dot{x} = p/m$ and $\dot{p}= -\nabla V$ where $H= p^2/2m + V(x)$.
Obviously, $div f =0$ which implies $\dot{\cal{S}}=0$. In particular this extends to the
 standard harmonic oscillator with $V(x)= (m\omega ^2/2) x^2$.

For the harmonic oscillator with friction, $\dot{x} = v$, $\dot{x}= -(\gamma /m)v - (\omega ^2/m) x$,
 we can adopt the observations of  subsection 3.1 with the two-by-two matrix $F$,
  whose first row contains only zeroes, while
$(F)_{21}= - \omega ^2/m$, $(F)_{22}= - \gamma /m$. Consequently  $tr F= -\gamma/m$.

A solution of the corresponding Liouville-type  equation  was  discussed in subsection 4.1.
The  Gibbs-Shannon entropy evolves in time according to  Eq.~(\ref{det}):
${\cal{S}}(t) =  {\cal{S}}(0) - {(\gamma t)/m}$ and ${\cal{S}} \rightarrow -\infty $ as
$t\rightarrow \infty  $.  Since  $\gamma >0$,  we
have   $\dot{\cal{S}} = - \gamma/m<0$.
There is no stationary density and hence no ${\cal{H}}_c(t)$.

An admixture of noise in the velocity/momentum rate equation   in  the  damped  harmonic oscillator case
 allows for the existence of a stationary density.
Let us consider, \cite{tyran,tyran1},  an example of the noisy damped harmonic
oscillator: $\dot{x}=p/m$, $\dot{p} = - (\gamma /m)p - (\omega ^2/m)p +\xi (t)$ where the
white noise term $\xi $  is normalized as follows $\left< \xi (t)
 \right> =0$, $\left<\xi (t)\xi (t')\right> = \sigma  \delta(t-t')$.
The corresponding  Fokker-Planck-Kramers equation for the probability density $f(x,v)$, with $v=p/m$ is:
\begin{equation}
{\frac{\partial f}{\partial t}} = - {\frac{\partial (vf)}{\partial x}} + {\frac{1}{m}}
{\frac{\partial [(\gamma v + \omega ^2 x)f]}{\partial v}} +
{\frac{\sigma ^2}{2m^2}} {\frac{\partial ^2f}{\partial v^2}}
\end{equation}
and has  a unique stationary solution:
\begin{equation}
f_*(x,v) = {\frac{\gamma \omega \sqrt{m}}{\pi \sigma ^2}} \exp \left[ -
{\frac{\gamma }{\sigma ^2}}(\omega ^2 x^2 + m v^2)\right]  \, . \label{stationary}
\end{equation}

A detailed, in part computer-assisted,  analysis of the temporal behavior of Gibbs-Shannon
and conditional K-L entropies evaluated for density  solutions of the above Kramers equation,
 with the initial data
\begin{equation}
f_0(x,v) = {\frac{1}{2\pi \sigma ^2_x \sigma ^2_v}} \exp\left( - {\frac{x^2}{2\sigma ^2_x}} -
 {\frac{v^2}{2\sigma ^2_v}}\right)
\end{equation}
has been made in Ref. \cite{tyran1}. We shall summarize  the  outcomes of this investigation .

In three basic regimes:  overdamped $\gamma ^2 > 4\omega ^2$, critical $\gamma ^2= 4\omega ^2$ and
underdamped $\gamma ^2 < 4\omega ^2$ cases, the conditional Kullback-Leibler  entropy
quantifies an approach of $f(x,v,t)$ towards $f_*(x,v)$  in terms of the  monotonic growth pattern
(this statement includes also  the case of $\dot{\cal{H}}_c(t)=0$).

The situation is entirely different, if we consider the Gibbs-Shannon entropy of $f(x,v,t)$.
Let us denote $\sigma _* = \sigma ^2/2\gamma \omega ^2$ and  $\alpha _x= \sigma ^2_x - \sigma _*$,
$\alpha _v = \sigma ^2_v - \omega ^2\sigma _*$.  The behavior  of ${\cal{S}}(t)$ sensitively  depends on the
mutual relations (signs, vanishing or non-vanishing of any or both etc.)
 between $\alpha _x$ and $\alpha _v$ and all details can be found in Ref.  \cite{tyran1}.

In the  overdamped  and critical  cases,  five  independent temporal behaviors are admitted. First
 three are of the monotonic
type, since $\dot{\cal{S}}$ is  vanishing, positive or negative. The fourth one admits a change of sign
of $\dot{\cal{S}}$ at certain $t_0 >0$ from  positive  to negative  plus the same scenario in reverse.
 The fifth temporal scenario shows  a passage through $\dot{\cal{S}}$-positive, negative and again
  positive stages of evolution plus  the  reverse (negative, positive, negative)  option.

The underdamped  case shows even  more intriguing patterns of temporal behavior. Namely, in addition to
the monotonic  negative or positive signs of $\dot{\cal{S}}$ we have also a conspicuous damped oscillation
of ${\cal{S}}(t)$,   where $\dot{\cal{S}}$ changes sign indefinitely, but an amplitude  of
oscillations performed by ${\cal{S}}(t)$ continually diminishes.

All these  \it diverse  \rm  temporal patterns are special  for the Gibbs-Shannon entropy. They  are
in turn  accompanied by a \it  unique  \rm  pattern of  the strictly
 monotonic growth  (or none) $\dot{\cal{H}}_c(t)\geq 0$  which is displayed by the conditional
  Kullback-Leibler entropy, \cite{tyran1}.

In close analogy with our considerations pertaining to the nontrivial power transfers between an
open dynamical  system and its thermal  environment, c.f. subsection 4.2, let us notice that the
invariant density Eq.~(\ref{stationary}) has the form analogous to this of $\rho _*$, Eq.~(\ref{inv}).
Indeed, we have:
\begin{equation}
f_*(x,v) = {\frac{1}{Z}}  \exp \left[ -
{\frac{2\gamma }{\sigma ^2}} E_{cl}(p,x) \right]
\end{equation}
with $1/Z=(\gamma \omega \sqrt{m})/(\pi \sigma ^2)$ and $E_{cl}(p,x) =  p^2/2m +  V(x)$ with
$V(x)=\omega ^2 x^2/2$ is an energy  of a classical harmonic oscillator at the $(x,p=mv)$ phase-space point.

Accordingly, we have:
\begin{equation}
{\cal{H}}_c(t) =  -  \int     f \,
\ln \left({\frac{f }{f _*}}\right )\, dx dv  =  {\cal{S}}(t) - \ln Z  -  {\frac{2\gamma }{\sigma ^2}}
     \langle  E_{cl}\rangle
\end{equation}
where  ${\cal{S}}(t)= - \int f \ln f dx dv$.
Therefore, it is an intrinsic property of our  dynamical system that
 $\dot{\cal{H}} = \dot{\cal{S}}  + \dot{\cal{Q}} \geq 0$,  where  we define
\begin{equation}
{\frac{d}{dt}} \langle  E_{cl}\rangle \doteq  - {\frac{\sigma ^2}{2\gamma }}  \dot{\cal{Q}}
\end{equation}
and clearly, $\dot{\cal{Q}}$  is  the direct analogue of the  previously introduced   power/heat
 transfer rate in the mean, c.f  , Eqs.~(\ref{heat})
and (\ref{heat1}).

\section{Conclusions}

 Standard notions of thermodynamical  entropy are  basically  used under  equilibrium or near-equilibrium
 conditions. The primary  built-in concept is an equilibrium (steady) state and the behavior  of entropy in
 the domain domain is seldom addressed.

If one attempts  to analyze  a dynamics  of an  approach towards  the prescribed steady  state,
it is necessary  to pass to the time domain where the   non-equilibrium  and  often rapid
dynamical processes  take place.
Various notions of entropy may be designed to quantify such non-equilibrium phenomena.

Our   analysis of simple diffusion-type  models indicates that  the  very  notion  of entropy, except perhaps
for the standard Clausius thermodynamical entropy,  is  non-universal and  purpose-dependent.
In particular,  the conditional  Kullback-Leibler   entropy is regarded (in reference to the "purpose")
 to be   the  only valid entropy growth  justification in terms of model systems, \cite{tyran,tyran1},
 (that  in conformity with the standard interpretation  of  the second law of thermodynamics
 for closed systems).

 However,  a deeper  insight into the underlying physical phenomena
  (power/heat  transfer processes in the mean)   is available only through  the differential (Gibbs-Shannon)
   entropy,    whose temporal   behavior is generically inconsistent with
   the "entropy growth" pattern.
   Moreover, the Gibbs-Shannon entropy balance equation   contains the
   conditional Kullback-Leibler entropy time rate as   an explicit non-negative  "entropy production"
   or rather "entropy accumulation" term, see e.g. subsections 4.2 and 4.3.
    The entropy dissipation may proceed through the previously mentioned
   mean power transfer mechanism, however the involved  "heat transfer"  expression  $\dot{\cal{Q}}$
    is not necessarily   positive-definite.

The  conditional   Kullback-Leibler  entropy is  an  appropriate  tool in case of "slow" processes, and
  in the asymptotic (large) time regime.
 The Gibbs-Shannon (differential, information) entropy  is perfectly suited for the
"shortest description length analysis", in particular for the study of
 rapid changes in time  of the probability distribution involved.

{\bf  Acknowledgement:} The paper has been supported by the Polish Ministry of Scientific Research and
Information Technology under the  (solicited) grant No PBZ-MIN-008/P03/2003.

\end{document}